\documentclass{PoSi}

\usepackage{verbatim}
\usepackage{amsmath}

\newcommand{\pbten}{$^{210}$Pb}

\newcommand{\ptwo}{\mbox{$^{32}$P}}
\newcommand{\sitwo}{\mbox{$^{32}$Si}}

\title{The DAMIC  dark matter experiment
}

\ShortTitle{The DAMIC  dark matter experiment
}

\author{\speaker{J.R.T.~de~Mello~Neto$^f$} on behalf of the DAMIC Collaboration}    
\author{
A.~Aguilar-Arevalo$^a$, 
D.~Amidei$^b$,
X.~Bertou$^c$,
D.~Bole$^b$,
M.~Butner$^{d, j}$,
G.~Cancelo$^d$,
A.~Casta\~{n}eda~V\'{a}zquez$^a$,
A.E.~Chavarria$^e$,
S.~Dixon$^e$,
J.C.~D'Olivo$^a$,
J.~Estrada$^d$, 
G.~Fernandez~Moroni$^d$,
K.P.~Hern\'{a}ndez~Torres$^a$,
F.~Izraelevitch$^d$,
A.~Kavner$^b$, 
B.~Kilminster$^g$,
I.~Lawson$^h$,
J.~Liao$^g$,
M.~L\'opez$^i$,
J.~Molina$^i$, 
G.~Moreno-Granados$^a$,
J.~Pena$^e$,
P.~Privitera$^e$,
Y.~Sarkis$^a$, 
V.~Scarpine$^d$,
T.~Schwarz$^b$,
M.~Sofo~Haro$^c$, 
J.~Tiffenberg$^d$,
D.~Torres~Machado$^f$,
F.~Trillaud$^a$,
X.~You$^f$ and
J.~Zhou$^e$\\
\llap{$^a$} Universidad Nacional Aut{\'o}noma de M{\'e}xico, M{\'e}xico D.F., M{\'e}xico \\  
\llap{$^b$} University of Michigan, Department of Physics, Ann Arbor, MI, United States \\  
\llap{$^c$} Centro At\'omico Bariloche - Instituto Balseiro, CNEA/CONICET, Argentina \\ 
\llap{$^d$} Fermi National Accelerator Laboratory, Batavia, IL, United States \\
\llap{$^e$} Kavli Institute for Cosmological Physics and The Enrico Fermi Institute, The University of Chicago, Chicago, IL, United States \\
\llap{$^f$} Universidade Federal do Rio de Janeiro, Instituto de  F\'{\i}sica, Rio de Janeiro, RJ, Brazil \\ 
\llap{$^g$} Universit{\"a}t Z{\"u}rich Physik Institut, Zurich, Switzerland \\   
\llap{$^h$} SNOLAB, Lively, ON, Canada \\  
\llap{$^i$}Facultad de Ingenier\'{\i}a - Universidad Nacional de Asunci\'on, Paraguay \\ 
\llap{$^j$} Northern Illinois University, DeKalb, IL, United States\\
 E-mail: \email{jtmn@if.ufrj.br
}}

\abstract{
The DAMIC (Dark Matter in CCDs) experiment uses high resistivity, scientific grade CCDs to search for dark matter. The CCD's low electronic noise  allows an unprecedently low energy threshold of a few tens of eV that make it possible to detect silicon recoils resulting from interactions of low mass WIMPs.  In addition the CCD's  high spatial resolution and the excellent energy response  results in very effective background identification techniques. The experiment has a unique sensitivity to dark matter particles with masses below 10 GeV/c$^2$.   Previous results have demonstrated  the potential of this technology, motivating the construction of DAMIC100, a 100 grams silicon target detector currently being installed at SNOLAB. 
In this contribution, the mode of operation and unique imaging capabilities of the CCDs, and how they may be exploited to characterize and suppress backgrounds will be discussed, as well as
physics results after one year of data taking. 
}

\FullConference{The 34th International Cosmic Ray Conference,\\
		30 July- 6 August, 2015\\
		The Hague, The Netherlands}

\begin{document}

\section{Introduction}

A well established body of evidence from astrophysics and cosmology supports the existence of cold dark matter as the major component of the material content of the universe.  The leading candidate for this dark matter is a hypothetical weakly interacting massive particle (WIMP) \cite{Jungman96,Bertone05}.  WIMPs could produce keV-energy nuclear recoils when scattering elastically off target nuclei in the detector.  Minimal supersymmetric extensions to the standard model favor particles above 50~GeV/c$^2$, while other models relating dark matter with the baryon asymmetry prefer masses around 5~GeV/c$^2$ \cite{Hamed09, Cohen10, Cheung09}.   Several experiments have reported statistically significant evidence of WIMPs scattering on light nuclear targets  \cite{Bernabei10, Agnese13}.

The DAMIC (Dark Matter in CCDs) experiment uses the bulk silicon of scientific-grade charge-coupled devices (CCDs) as the target for coherent WIMP-nucleus elastic scattering. Due to the low readout noise of the CCDs and the relatively low mass of the silicon nucleus,  CCDs are ideal instruments for the identification of the nuclear recoils with keV-scale energies and lower from WIMPs with masses < 10 GeV/c$^2$.  

The first DAMIC  measurements were performed in a shallow underground site at Fermilab using several 1-gram CCD detectors developed for the Dark Energy Survey (DES) camera (DECam) \cite{Flaugher12}. With 21g-days DAMIC produced the best upper limits on the cross-section for WIMPs below 4 GeV/c$^2$ \cite{Barreto12}. DAMIC is now located in SNOLAB laboratory 2 km below the surface in the Vale Creighton Mine near Sudbury, Ontario, Canada.

\section{The DAMIC detectors}

\begin{figure}
    \centering{ \includegraphics[width= 1.0\textwidth]{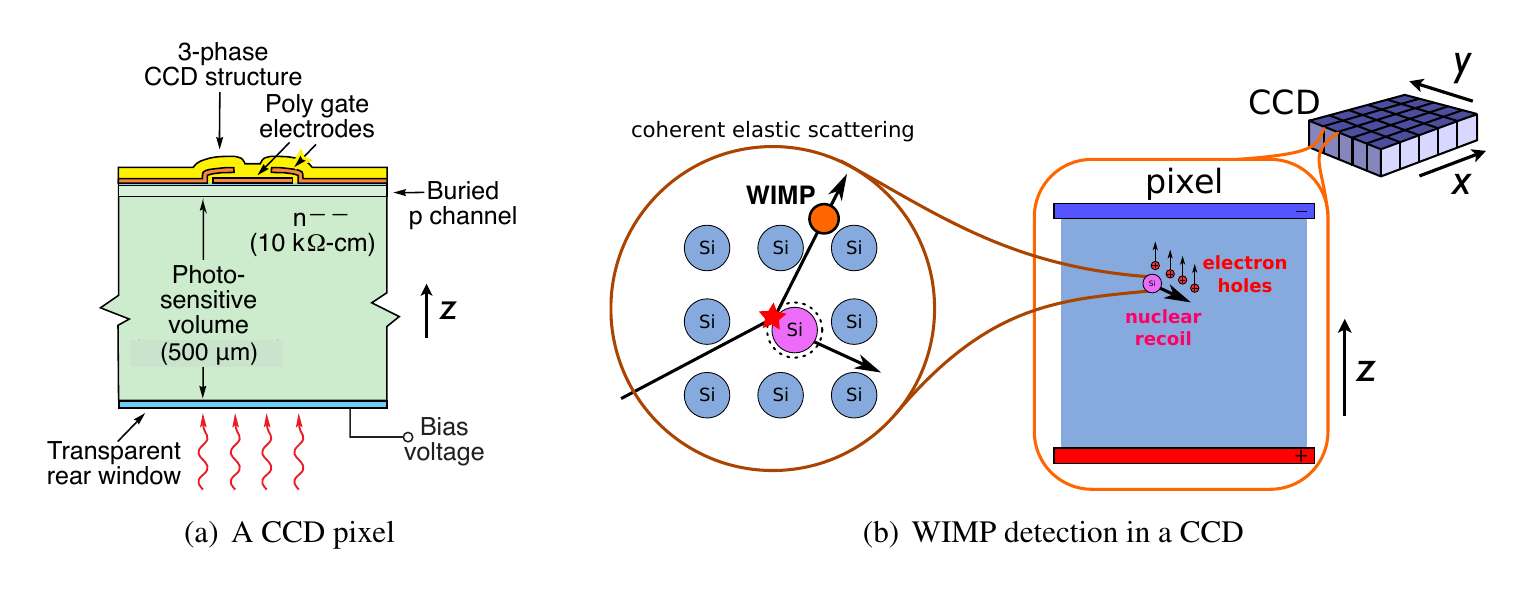}}
     \caption{a) Cross-sectional diagram of a  15 $\mu$m $\times$ 15 $\mu$m pixel in a fully depleted, back-illuminated CCD. The thickness of the gate structure and the backside ohmic contact are $\le 2 \mu$m. The transparent rear window has been eliminated in the DAMIC CCDs. b) Dark matter detection in a CCD. 
A WIMP scatters with a silicon nucleus in the active region, producing ionization from the nuclear recoil which drifts along the z-direction and is collected at the CCD gates.  }
     \label{ccd}
     \end{figure}

The DAMIC CCDs feature a three-phase polysilicon gate structure with a buried p-channel. The CCDs are typically 8 or 16 Mpixels, with pixel size of 15 $\mu$m $\times$ 15 $\mu$m, with a total surface area of tens of cm$^2$.
 The CCDs are  675 $\mu$m thick, for a  mass up to 5.2 g. A   high-resistivity (10-20 k$\Omega\,$cm) n-type silicon  allows  for a low donor density in the substrate ($\sim 10^{11}$ cm$^{-3}$), which leads to fully depleted operation at low values of the applied bias voltage ($\sim $40 V for a 675 $\mu$m-thick CCD).    Fig. \ref{ccd} shows a cross-sectional diagram of a CCD pixel, together with a sketch depicting the WIMP detection principle.  The substrate voltage also controls the level of lateral diffusion of the charge carriers as they drift the thickness of the
 CCD.  The lateral spread (width) of the charge recorded on the CCD $x$-$y$ plane may be used to reconstruct the $z$-coordinate of a point-like interaction \cite{Chavarria15}.

Long exposures are taken in DAMIC ($\sim$8 hours) in order to minimize the number of readouts and consequently the number of pixels above a given threshold due to readout noise fluctuations.  The CCD dark current due to thermal excitations  (< 0.1 $e^-$pix$^{-1}$day$^{-1}$ at the operating temperature of $\sim$140 K) contributes negligibly to the noise.  During readout, the charge held at the CCD gates is measured by shifting charge row-by-row and column-by-column via phased potential wells  to a low capacitance output gate.  The inefficiency  of charge transfer from pixel to pixel is as low as $10^{-6}$.  The readout noise for the charge collected in a pixel is $\sim$2 $e^-$ which corresponds to an uncertainty of $\sim$7 eV of ionizing energy in Silicon. 

 \begin{figure}
   \centering{  \includegraphics[width=1.0\textwidth]{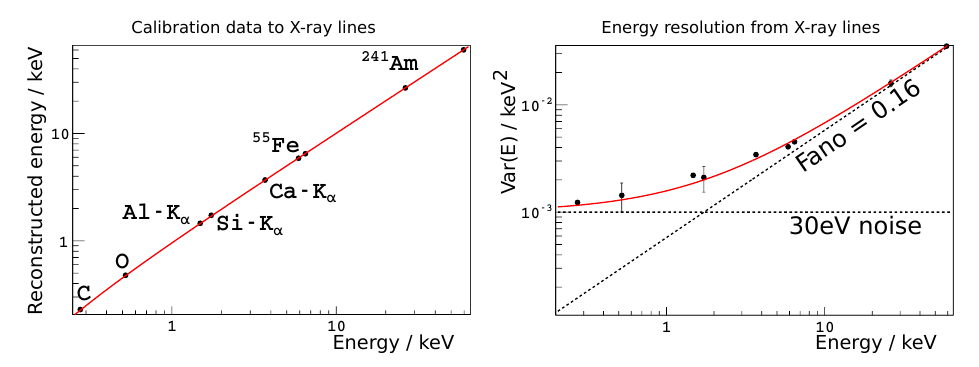}}
     \caption{a)  Reconstructed energy of an X-ray line compared to is true energy.  The labeled K$_{\alpha}$ markers are fluorescence lines from elements in the Kapton target and other materials in the CCD setup.  The $^{55}$Fe and $^{241}$Am markers are X-rays emitted by the radioactive sources. Linearity in the measurement of ionization energy is demonstrated from 0.3 kev to 60 keV.   b)  Variance of the X-ray lines as a function of energy. The effective Fano factor is 0.16, typical for a CCD \cite{Janesick01}. }
     \label{calibration}
     \end{figure}
     
Calibrations with a $^{55}$Fe source, with fluorescence X-rays from a Kapton target exposed to the $^{55}$Fe source and with  $\alpha$s from $^{241}$Am were performed. 
As shown in  fig.~\ref{calibration},  the detectors present an excellent linearity and energy resolution (55 eV RMS at 5.9 KeV) for electron-induced ionization, as measured with X-ray sources \cite{Chavarria15}.
The ionization efficiency of nuclear recoils is significantly different than that of electrons. Previous measurements have been done down to energies of \mbox{3-4~keV$_r$} \cite{Dougherty92, Gerbier90} in agreement with Lindhard theory  \cite{Ziegler85}.  From this, DAMIC's nominal 50~eV$_{ee}$ threshold corresponds to $\sim$0.5~keV$_r$.

\begin{figure}
   \centering{  \includegraphics[width=1.0\textwidth]{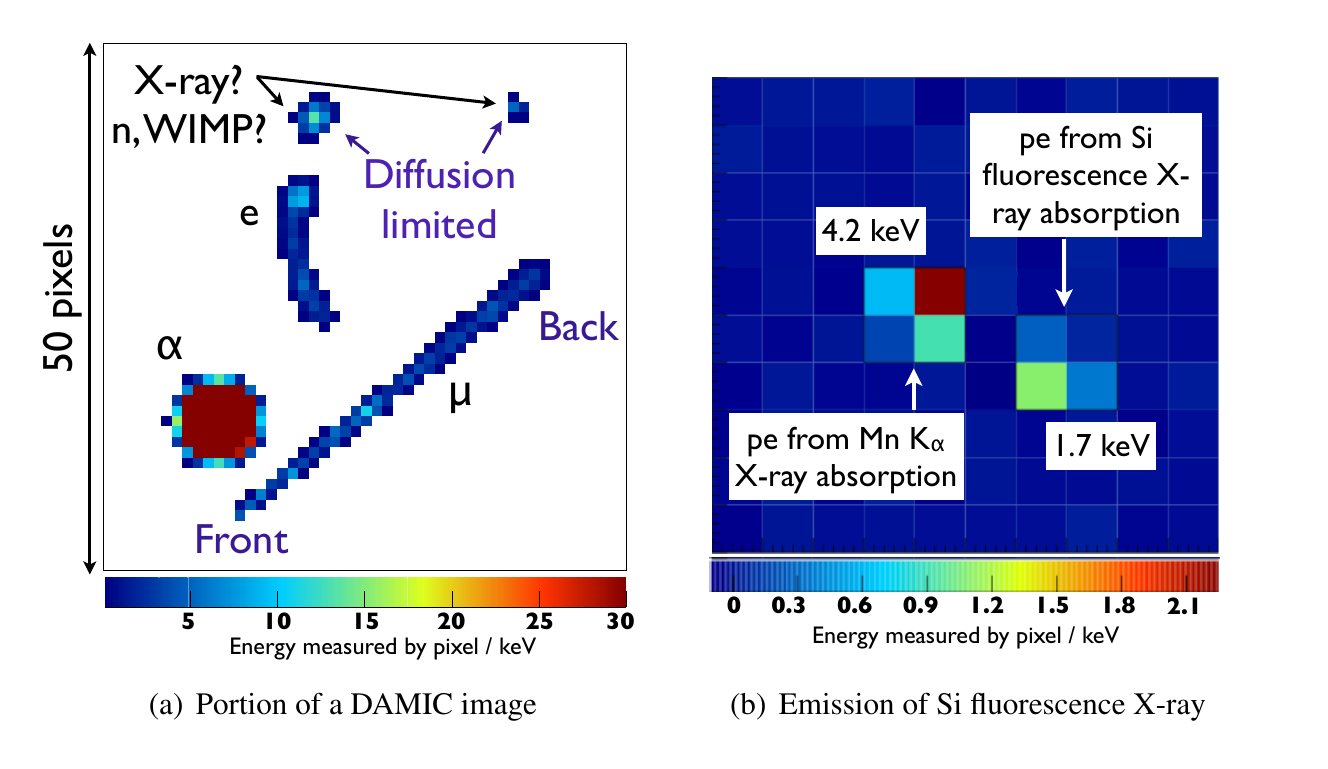}}
     \caption{a) 50$\times$50 pixel segment of a DAMIC image  exposed to a $^{252}$Cf source when the detector was at ground level. Only pixels with deposited energy $ > $0.1keV$_{ee}$ are colored. b) Event with two nearby clusters detected after illuminating the CCD with a $^{55}$Fe source.  The 1.7 keV cluster is a photoelectron (pe) from the absorption of a Si fluorescence X-ray, emitted following photoelectric absorption of the incident 5.9 keV Mn K$_{\alpha}$ X-ray in a nearby site.  }
     \label{images}
     \end{figure} 

The total charge and shape of each hit is extracted using dedicated image analysis tools.  In fig.~\ref{images} a sample of tracks recorded during a short exposure at sea level to a $^{252}$Cf source is shown.    Clusters from different types of particles may be observed. Low energy electrons and nuclear recoils, whose physical track length is <15 $\mu$m, produce  ''diffusion limited''  clusters, where the spatial extent of the cluster is dominated by charge diffusion. Higher energy electrons ($e$), from either Compton scattering or $\beta$ decay, lead to extended tracks. $\alpha$  particles in the bulk or from the back of the CCD produce large round structures due to the plasma effect \cite{Estrada11}. Cosmic muons ($\mu$) pierce through the CCD, leaving a straight track. The orientation of the track is immediately evident from its width, the end-point of the track that is on the back of the CCD is much wider than the end-point at the front due to charge diffusion.

\section{The DAMIC experiment at SNOLAB}
 \begin{figure}
   \centering{  \includegraphics[width=1.0\textwidth]{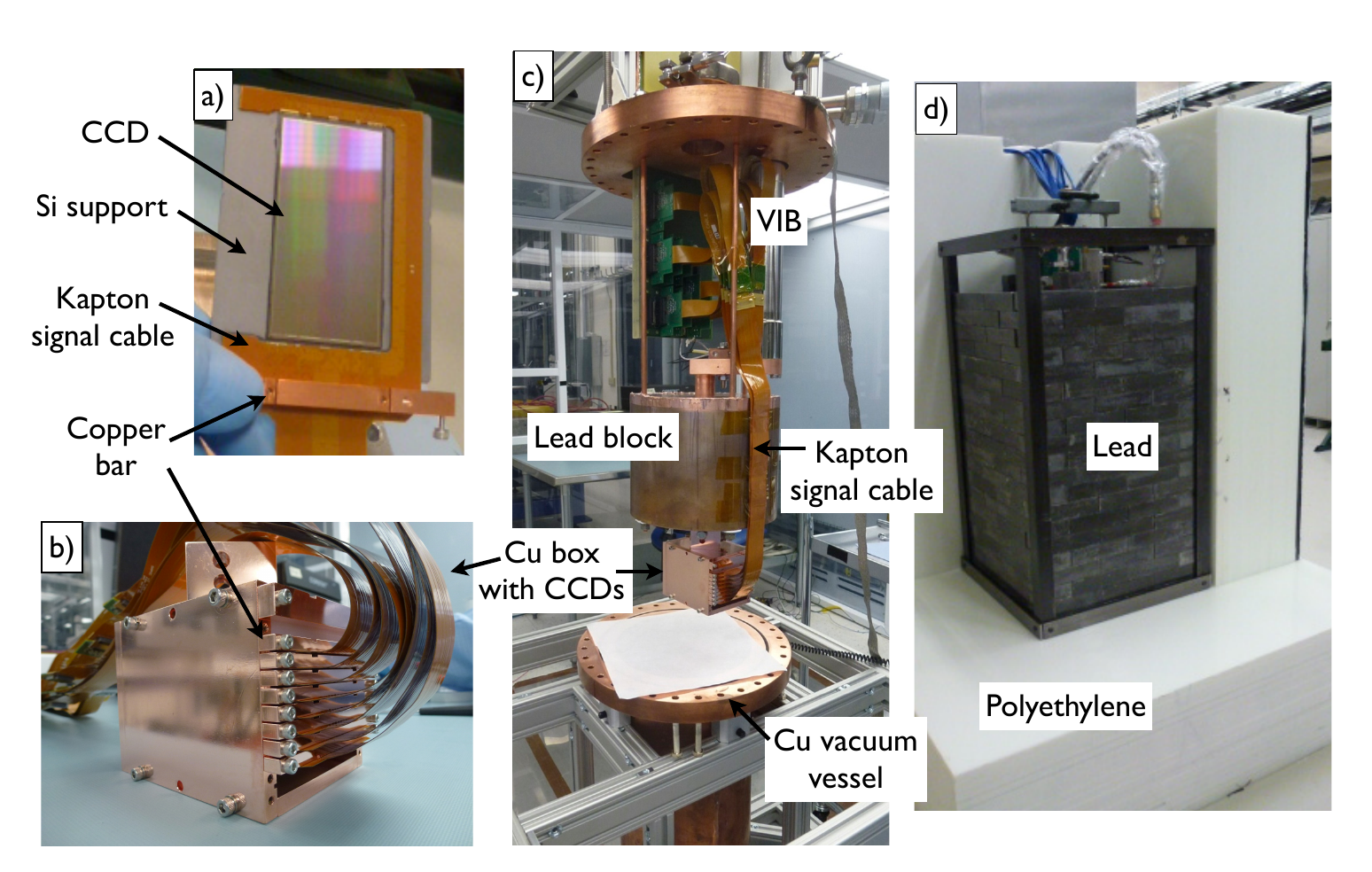}}
     \caption{a) A packaged DAMIC CCD. b) The copper box housing the CCDs. c) Components of the DAMIC setup, ready to be inserted in the vacuum vessel. d) The vessel inside the lead castle, during installation of the polyethylene shield.}
     \label{setup}
     \end{figure}
\noindent Fig.~\ref{setup}  shows the infrastructure already installed in SNOLAB.  A packaged CCD (2k$\times$4k, 8 Mpixel, 500 $\mu$m-thick) is shown in fig \ref{setup}a. The device is epoxied to a high-purity silicon support piece. The Kapton signal flex cable bring the signals from the CCDs up to the vacuum interface board (VIB).    The cable is also glued to the silicon support. A copper bar facilitates the handling of the packaged CCD and its insertion into a slot of an electropolished copper box (fig~\ref{setup}b). The box is cooled to $\sim$140 K inside a copper vacuum vessel ($\sim10^{-6}$ mbar). An 18 cm-thick lead block hanging from the vessel-flange shields the CCDs from radiation produced by the VIB, also located inside the vessel  (fig~\ref{setup}c). The CCDs are connected to the VIB through Kapton flex cables, which run along the side of the lead block. The processed signals then proceed to the data acquisition electronic boards. The vacuum vessel is inserted in a lead castle (fig \ref{setup}b) with 21 cm thickness to shield the CCDs from ambient  $\gamma$-rays. The innermost inch of lead comes from an ancient Spanish galleon and has negligible $^{210}$Pb content, strongly suppressing the background from bremsstrahlung $\gamma$s produced by $^{210}$Bi decays in the outer lead shield. A 42 cm-thick polyethylene shielding is used to moderate and absorb environmental neutrons.

\section{Measurements of radioactive contamination}
The ultimate sensitivity of the experiment is determined by the rate of the radioactive background that mimics the nuclear recoil signal from the WIMPS.  The SNOLAB underground laboratory has low intrinsic background due to its 6000 m.w.e. overburden.  Dedicated screening and selection of detector shielding materials, as well as radon-suppression methods, are extensively employed to decrease the background from radioactive decays in the surrounding environment.  The measurement of the intrinsic contamination of the detector is fundamental.  For silicon-based experiments the cosmogenic isotope $^{32}$Si, which could be present in the active target,  is particularly relevant since its $\beta$ decay spectrum extends to the lowest energies and may become an irreducible background.  The analysis methods used to establish the contamination levels exploit the unique spatial resolution of the CCDs. 
 \begin{figure}
   \centering{  \includegraphics[width=1.0\textwidth]{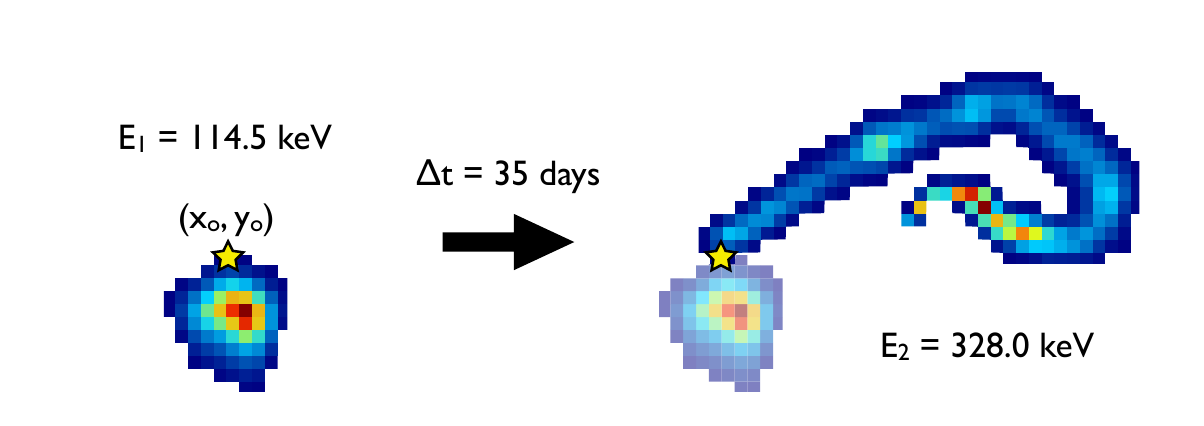}}
     \caption{Candidate $\beta$ decay sequence found in data. The first cluster had 114.5 keV of energy.  A second cluster, with energy 328.0 keV, was observed in an image taken 35 days later. Both tracks appear to originate from the same point (yellow star) in the CCD $x$-$y$ plane. }
     \label{beta}
     \end{figure}
     
The identification of $\alpha$-induced clusters is the first step in establishing limits on uranium and thorium contamination  \cite{Aguilar-Arevalo15}.  
Radiogenic $\alpha$s lose most of their energy by ionization, creating a dense column of electron-hole pairs that satisfy the plasma condition \cite{Estrada11}.  
For interactions deep in the substrate, the charge carriers diffuse laterally and lead to round clusters of hundreds of micrometers in diameter, whereas  $\alpha$ particles that strike the front of the CCD lead to mostly vertical clusters according to a phenomenon known as ``blooming'' \cite{Janesick01}. 
    Simple criteria are sufficient to efficiently select and classify $\alpha$s. Spectroscopy of plasma $\alpha$s can be used to establish limits on $^{210}$Pb,  $^{238}$U and $^{232}$Th contamination in the bulk of the CCD. 
 In special DAMIC runs, with a dynamic range optimized for $\alpha$ energies, four plasma $\alpha$s whose energies are consistent with $^{210}$Po were observed. One of them cannot be $^{210}$Po, as it coincides spatially with two higher energy $\alpha$s recorded in different CCD exposures, and is therefore likely part of a decay sequence. 
  When interpreting the other three as bulk contamination of $^{210}$Po (or  $^{210}$Pb), an upper limit of < 37 kg$^{-1}$d$^{-1}$ (95\% CL) is derived.  In the  $^{238}$U chain, the isotopes $^{234}$U, $^{230}$Th and $^{226}$Ra decay by emission of $\alpha$s with energies 4.7-4.8 MeV.  
  Since the isotopes' lifetimes are much longer than the CCD exposure time, their decays are expected to be uncorrelated. No plasma $\alpha$s were observed in the 4.5-5.0 MeV energy range, and an upper limit on the $^{238}$U contamination of < 5  kg$^{-1}$d$^{-1}$ (95\% CL) is correspondingly derived (secular equilibrium of the isotopes with $^{238}$U was assumed).   A similar analysis results in a upper limit of < 15 kg$^{-1}$d$^{-1}$ (95\% CL) on $^{232}$Th contamination in the CCD bulk \cite{Aguilar-Arevalo15}.
 
A search for decay sequences of two $\beta$ tracks was performed to identify radioactive contamination from $^{32}$Si and $^{210}$Pb and their daughters.   $^{32}$Si leads to the following decay sequence: 
  \begin{align*}
\mbox{\sitwo} &\longrightarrow \mbox{\ptwo} +\beta^-  ~\rm{with } ~\tau_{1/2} = \rm{150\,y, ~Q-value=227\,keV}\\
\mbox{\ptwo} &\longrightarrow \mbox{$^{32}$S} +\beta^- ~ \rm{with} ~\tau_{1/2} = \rm{14\,d, ~Q-value = 1.71\,MeV}
\end{align*}
A total of  13 candidate pairs were observed in the data. With detailed Monte Carlo simulations the overall efficiency for detection of   \sitwo\,--\ptwo\ decay sequences in the data set was determined to be $\epsilon_{\rm Si} = 49.2$\%.  The number of accidental pairs was also determined with simulations.  The decay rate was estimated to be $80^{+110}_{-65}$\,kg$^{-1}$\,d$^{-1}$ (95\% CI) for \sitwo\ in the CCD bulk \cite{Aguilar-Arevalo15}.  With a similar procedure the upper limit on the \pbten\ decay rate in the CCD bulk has been deduced as  $<$33\,kg$^{-1}$\,d$^{-1}$ (95\% CL).

\section{Dark matter search}
 \begin{figure}
    \centering{ \includegraphics[width=0.85\textwidth]{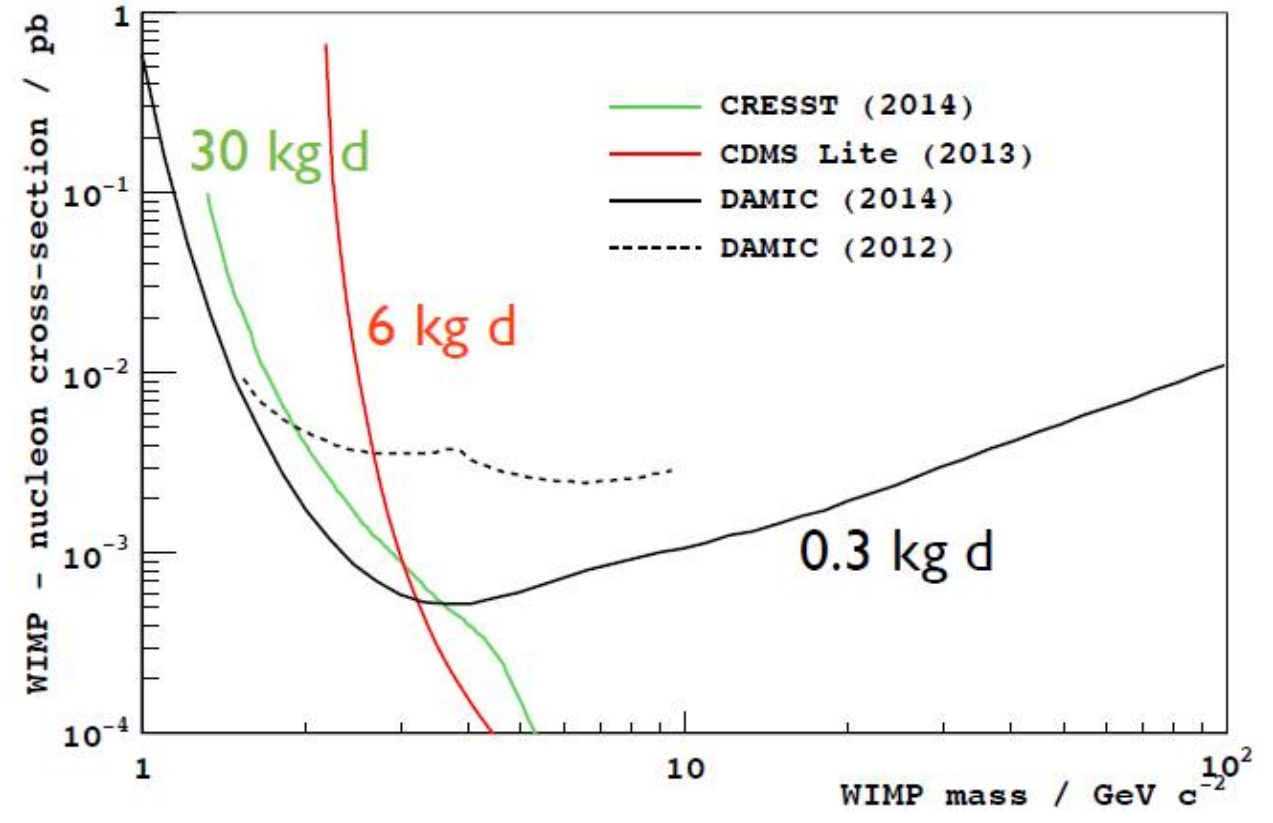}}
     \caption{Cross section exclusion limit at 90\% CL for the DAMIC 2014 results (solid black) compared to DAMIC 2012 (dashed black) \cite{Barreto12},  CRESST 2014 (solid green) \cite{CRESST14} , CDMSlite 2013 (solid red) \cite{CDMS13}.}
     \label{limits}
     \end{figure}
The data acquired in 2014 came from two CCDs  500 $\mu$m   thick and 2.2 g exposed for 36 days  and another    CCD 675 $\mu$m thick and 2.9 g exposed for 7 days.  We assumed a local WIMP density of 0.3 GeV/cm$^3$, dispersion velocity for the halo  of 220 km/s, earth velocity of 232 km/s and a escape velocity of 544 km/s.  The Lindhard model was used to obtain recoil energies as discussed above. The data analysis proceeded with a two-dimensional gaussian fit to each hit in the images.  The noise was used to set the signal threshold and simulation was used to estimate the efficiency down to the threshold. Based on this efficiency, the total exposure was calculated as $\sim$0.3 kg-d. The recoil spectrum was fitted with the described  WIMP model and no candidates were found.  The resulting 90\% CL are show in fig. \ref{limits}, together with CRESST and CDMSlite results.  The DAMIC results constitute the new best limits for dark matter particles of masses below 3 GeV/c$^2$.

\section{Conclusions}
We have shown that DAMIC is producing high quality science and it is a leading experiment at low WIMP mass.  The CCD detectors with their unique imaging capabilities allow us to measure internal contamination of silicon in a unique way. Stringent 95\% CL upper limits on the presence of radioactive contaminants in the silicon bulk were placed. 
The dark matter search will be performed with an upgraded experiment,  DAMIC100, a  low background detector with 100 g of sensitive mass, consisting of 18 CCDs, each of them with  16 Mpix, 675 $\mu$m thickness  and 5.5 g. The measured levels of radioactive contamination are already low enough for its successful operation. We are currently testing  the new CCDs. The baseline plan is to operate the DAMIC-100 experiment for one year to collect approximately 30 kg-day exposure by the end of 2016.

\end{document}